\documentclass[pre,floatfix,showpacs,preprint]{revtex4}

\usepackage{amsmath}
\usepackage{amssymb}
\usepackage{graphicx}
\usepackage{color}
\usepackage{psfrag}
\usepackage{verbatim}
\usepackage[tight,sf]{subfigure}

\usepackage{ulem}
\newcommand{\romd}{{\text{d}}}

\newcommand{\sfR}{{\textsf{R}}}

\newcommand{\VECi}{{\boldsymbol{i}}}
\newcommand{\VECj}{{\boldsymbol{j}}}

\newcommand{\VECn}{{\boldsymbol{n}}}

\newcommand{\VECt}{{\boldsymbol{t}}}
\newcommand{\VECu}{{\boldsymbol{u}}}
\newcommand{\VECx}{{\boldsymbol{x}}}
\newcommand{\VECy}{{\boldsymbol{y}}}
\newcommand{\VECz}{{\boldsymbol{z}}}

\newcommand{\VECD}{{\boldsymbol{D}}}

\newcommand{\VECM}{{\boldsymbol{M}}}

\newcommand{\VECX}{{\boldsymbol{X}}}

\newcommand{\CALE}{{\mathcal{E}}}

\newcommand{\CALK}{{\mathcal{K}}}

\newcommand{\ellipticK}[1]{\CALK\left[ #1\right]}

\newcommand{\ellipticE}[1]{\CALE [ #1 ]}
\newcommand{\ellipticpiadjust}[3]{\operatorname{\Pi}\left[  #1,#2,#3 \right]}
\newcommand{\ellipticpicompleteadjust}[2]{\operatorname{\Pi}\left[  #1,#2 \right]}
\newcommand{\Jacobisnadjust}[2]{\operatorname{sn}\left[  #1,#2 \right]}
\newcommand{\Jacobisnsquare}[2]{\operatorname{sn}^{2}[ #1,#2 ]}

\newcommand{\Jacobiamadjust}[2]{\operatorname{am}\left[  #1,#2 \right]}

\newcommand{\RR}{\mathbb{R}}

\newcommand{\se}{\Psi}

\newcommand{\Cpar}{C}

\usepackage{mathrsfs}

\begin{document}
\title{Dipoles in thin sheets}

\author{Jemal Guven}
\affiliation{Instituto de Ciencias Nucleares, %
             Universidad Nacional Aut\'onoma de M\'exico; Apdo.\ Postal 70-543, 04510 M\'exico D.F., Mexico}
\author{J. A. Hanna}
\affiliation{Department of Physics, University of Massachusetts;  666 N.\ Pleasant St.,  Amherst, MA 01003, USA}
\author{Osman Kahraman}
\affiliation{Equipe BioPhysStat, ICPMB-FR CNRS 2843, Universit\'{e} de Lorraine; 
            1, boulevard Arago, 57070 Metz,
            France}
\author{Martin Michael M\"uller}
\affiliation{Equipe BioPhysStat, ICPMB-FR CNRS 2843, Universit\'{e} de Lorraine; 
            1, boulevard Arago, 57070 Metz,  France}
\affiliation{CNRS, Institut Charles Sadron; 23 rue du Loess BP 84047, 67034 Strasbourg, France}

\date{\today}
\begin{abstract}
A flat elastic sheet may contain pointlike conical singularities that carry a metrical ``charge'' of Gaussian curvature. 
Adding such elementary defects to a sheet allows one to make many shapes, in a manner broadly 
analogous to the familiar multipole construction in electrostatics. However, here the underlying field theory is non-linear, 
and superposition of intrinsic defects is non-trivial as it must respect the immersion of the resulting surface in three 
dimensions.  We consider a ``charge-neutral'' dipole composed of two conical singularities of opposite sign.
Unlike the relatively simple electrostatic case, here there are two distinct stable minima and an infinity of unstable equilibria.  
We determine the shapes of the minima and evaluate their energies in the thin-sheet regime where bending dominates 
over stretching. Our predictions are in surprisingly good agreement with experiments on paper sheets.
\end{abstract}

\pacs{68.55.-a,46.32.+x,02.40.Hw}

\maketitle


\bibliographystyle{plain}


\section{Introduction}

The first step towards an understanding of a linear field theory is often to examine its fundamental singular solutions.  These serve as elementary components of complicated fields constructed \emph{via} linear superposition; familiar examples abound in electrostatics \cite{Jackson75} and viscous \cite{KimKarrila05} and inviscid \cite{Saffman92} fluid flows.

There is currently no analogous procedure for generating the shape of an elastic sheet from singularities in its Gaussian curvature.  The corresponding field equations are non-linear, and the canonical singularities may be more exotic than points or lines \cite{Witten07,AmirbHearle86,ChaiebMelo99,Liang2005,Liang2008}. Yet in a thin sheet, the simplest defects are pointlike and conical, and such a defect bears a strong resemblance to a monopole.
The intrinsic geometry of a conical monopole can be characterized completely by an angle $\Psi$.  If we remove a wedge of material ($\Psi<0$),
the only bending equilibrium of an isolated sheet is a circular ice cream cone. If this deficit is
replaced by a surplus ($\Psi>0$), an infinite number of dihedrally symmetric equilibrium shapes exists \cite{short}, but the two-fold symmetric state is the only stable one \cite{JemalMartinPablo2012}.

\begin{figure}
\centering
\includegraphics[height=5cm]{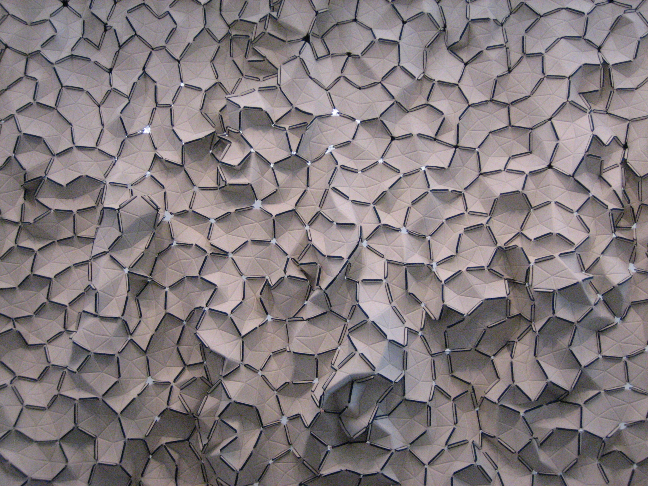}
\hspace{0.5cm}
\includegraphics[height=5cm]{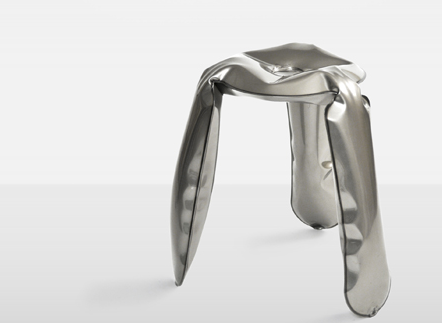}
\caption[]{Surface constructions riddled with pointlike singularities. 
Left: closeup of R Bouroullec and E Bouroullec ``Clouds'' (2008). 
Right: Zieta Prozessdesign ``Plopp'' stool (2009).}
\label{fig:art}
\end{figure}

The question of how to construct multipolar analogues of a conical defect is an obvious next step, one that has already been hinted at in the treatment of disclination dipoles as ``dislocations'' in crystalline membranes \cite{SeungNelson88}. 
Composite defects can always be constructed by placing additional pointlike defects on the sheet (Fig.~\ref{fig:art}). When a cardboard box is folded from a flat template,
positive Gaussian curvature is generated at its eight vertices.  A general defect will consist of a combination of deficits and surpluses.
The simplest composite defect is a dipole.  Defects like these can arise during metric-preserving deformations of flat sheets by various types of forcing \cite{ChaiebMelo97,MoraBoudaoud02,Hamm04,Das07,Walsh11} 
or through reconstructive surgery of a surface. More complicated elastic defects may arise through organismal growth \cite{JulienMartine,Mandoli1998}, chemical swelling \cite{Sharon2007,Kim2012}, or the formation of coherent precipitates (Eshelby inclusions) in crystalline alloys \cite{Eshelby}.

We consider the extrinsic geometry of the analogue of a pure electrostatic dipole, a deficit and a surplus separated by a finite distance, carrying a total Gaussian curvature ``charge'' of zero.
The distance, and charges, must be finite as we have already hit a snag in our construction:  the classical dipole limit is inaccessible as we are unable to remove an arbitrarily large wedge from a plane.  A greater complication arises from the absence of linear superposition. It is not generally possible to sew two cones
together without introducing stretching somewhere on the surface. One resolution, which we will use in this article, is to approximate the geometry by two conical parts joined by a ridge, with planar faces on either side interpolating between the cones. 

We will begin by describing this geometry. The energy associated with the sheet is bending energy, quadratic in the mean curvature. 
We will ignore the energy associated with the formation of the ridge between the two defects. 
We will see that--- unlikely as this may appear--- this is legitimate for understanding the broad features of a dipole. Even this simple model displays behavior without any electrostatic analogues. There are two distinct stable dipole states with different energies, separated by a high energy barrier; the ground state is quite mechanically rigid.


\section{Conformations of the dipole}
\begin{figure}
  \subfigure[]{\includegraphics[width=0.502\textwidth]{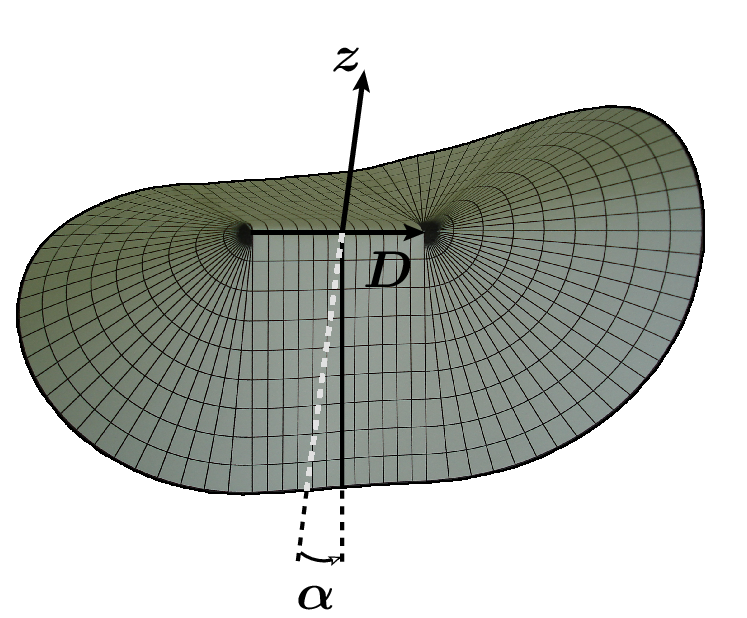}\label{fig:torquedipole}}
  \hspace*{0.5cm}
  \subfigure[]{\raisebox{.65cm}{\includegraphics[width=0.452\textwidth]{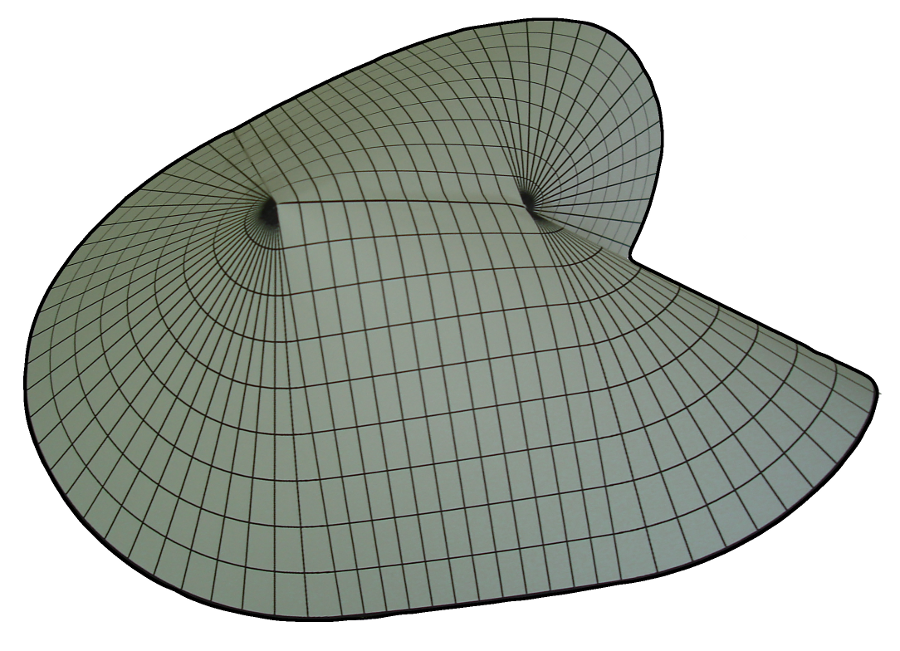}}\label{fig:forcedipole}}
  \caption[]{Dipole geometries modeled with an A4 sheet of paper (photographs). (a) Torque dipole and (b) force dipole of strength $\Psi=\frac{\pi}{4}$ and dipole length $D=6\,$cm.
}
\label{fig:epoles}
\end{figure}

A simple way to familiarize oneself with the peculiarities of a dipole in an unstretchable sheet 
is to tinker with a piece of paper \cite{WittenLi93}.  Take two points near the center of a rectangular sheet and connect them with a line parallel to the longer side. To construct a dipole of strength $\Psi$, first cut out a wedge with an angle $\Psi$ at one of the two points, 
join the two exposed sides in the sheet, and round off the edges.  Then, make a cut from the second point to the boundary, insert the wedge, and round off again.   

Two different stable shapes are possible (Fig.~\ref{fig:epoles}): either the two sides extend into two opposite regions of space, 
or the whole sheet lies beneath the line joining the two points. 
The two configurations are stable with respect to small deformations, but can be transformed into one another if significant forces are applied.  Outside the singularities, the dipole in Fig.~\ref{fig:forcedipole} resembles the zero-deficit ``developable cone'' obtained, among other ways, by depressing a sheet into a cup with 
a point force at its center \cite{benpom,ChaiebMelo97,Chaieb98,cerdamaha}. 
The second dipole, in Fig.~\ref{fig:torquedipole}, seems to represent the analogous application of a point torque.
Accordingly, we will refer to the two configurations as \emph{force} and \emph{torque} dipoles.\footnote{Please note that this labeling is an analogy and not a precise statement about the system.}

In both cases, the surface can be approximated by planar parts which connect a deficit to a surplus cone.
The two planar regions of the torque dipole appear to form triangles extending  to the boundary of the sheet.  The two triangles meet along a straight ridge, with dihedral angle $2\alpha$, connecting the conical singularities.
The force dipole can be approximated in similar fashion, but one must exercise a little more care: in the paper model, the ridge is not straight, and it is not 
completely clear how the flat regions can be defined. We will return to this point below. For the theoretical model, we will assume that the dipole consists of a pair of unstretchable cones joined by a straight ridge and planar regions. The dipole vector $\VECD$ points from the deficit singularity to the surplus singularity along the ridge.  This vector, together with the apparent mirror plane of the equilibrium dipole geometries, defines the perpendicular vertical ${\VECz}$.


\section{Dissecting the problem}
To construct the global geometries,  we first consider the two conical defects separately.  We will use the subscript $\pm$ in the following only when the distinction between the two cones is relevant.
Each of the cones can be described in terms of an open curve $\Gamma_\pm:s\to\VECu_\pm(s)$ on the unit sphere, where the arc length $s$ runs from 0 to $L_\pm$. 
The value of $L_\pm$ depends on the dipole strength $\se$ and, implicitly via the coupling between the two 
cones, on the dihedral angle $2\alpha$. 
Each conical surface is given by the vector function
$\VECX_\pm(r,s)=r\VECu_\pm(s)\in\RR^{3}$, where $r$ is the radial distance to the apex. 
The two tangent vectors to each cone, $\VECu$ and $\VECt=\partial_s\VECu=\VECu'$,
form a right-handed orthonormal surface basis together with the normal $\VECn=\VECu\times\VECt$. 
Whereas the surface is flat in the direction of $\VECu$, its curvature along $s$ is given by $\kappa=-\VECn\cdot\VECt'$. 
The latter corresponds to the geodesic curvature of $\Gamma$ on the unit sphere and should not be mistaken for the Frenet curvature.

We will have to identify the configurations that minimize the bending energy of an open unstretchable
cone of radius $R$ with surplus or deficit angle. Since tangent vectors and curvatures must be continuous 
across the boundaries between flat and conical parts, $\kappa$ must vanish along 
two fixed radial lines. If we cut off the apex at $r_0$ and integrate over the radial direction, the bending energy is given by
$B = (a/2) \int_{\Gamma} \romd s \, \kappa^{2}$, where $a=\ln{(R/r_{0})}$. 
The unstretchability of the sheet can be implemented by adding a constraint term to the energy
functional which fixes the metric \emph{via} a set of Lagrange multiplier functions $T^{ab}$. These can be identified with a conserved 
tangential stress \cite{paperfolding}, which for the present case is encoded in a constant $\Cpar$. For the conical sheet, $\{a,b\}\in\{r , s\}$, $T^{ss}=-T^{rr}/r^2=-\Cpar /r^4$, and $T^{rs} =0$.

For an isolated conical monopole, there is a remarkably simple way to determine the shape of such a surface \cite{paperfolding,short}.  Recall that one can find 
a conserved Noether current for every continuous symmetry of a system. The fixed apex breaks translational invariance of the monopole, but
the bending energy remains rotationally invariant. The corresponding conserved vector $\VECM$ is given by 
$  \VECM = \left( \kappa^{2}/2 - \Cpar \right) \, \VECn + \kappa ' \VECt + \kappa \VECu$. 
Its square yields the first integral of the shape equation
of the monopole:
\begin{equation}  \label{eq:firstintegral}
  M^{2} - \Cpar^{2}
  = \kappa'^{2} + \kappa^{4}/4 + (1 - \Cpar) \kappa^{2}
  \; ,
\end{equation}
analogous to planar Euler \emph{elastica}.  

The problem we consider here is more complicated than that of an isolated monopole: the surface is modeled by combining two individual cones, neither of which forms a closed surface.  One might be skeptical as to whether it is correct to appeal to the rotational invariance underlying conservation of the torque vector $\VECM$ for each individual cone. However, we will see below that this yields a valid approximation for the
constituent cones of the dipoles.  The interaction between the two cones will occur entirely through boundary conditions.

Eqn.~(\ref{eq:firstintegral}) can be solved for $\kappa$ in terms of elliptic functions. Since $\kappa$ vanishes on the boundaries, one can deduce directly that 
the constant of integration $M^{2} - \Cpar^{2}$ has to be positive. One obtains for the curvature (see appendix~\ref{app:determinecurvature}): 
$\kappa_\pm(s) =  \mp 4 n_\pm \sqrt{-k}\, (\ellipticK{k}/L_\pm) \;\Jacobisnadjust{2 n_\pm (\ellipticK{k}/L_\pm) \, s}{k}$, 
where 
$\Jacobisnadjust{s}{k}$ is the sine of the Jacobi amplitude $\Jacobiamadjust{s}{k}$ with parameter $k\le 0$, 
$\ellipticK{k}$ is the complete elliptic integral of the first kind \cite{Abramowitz}, and $n_\pm$ denotes the number of lobes of each cone.  In the paper model, $n_-=1$ and $n_+=3$. 
%
%
\begin{figure}
\centering
  \subfigure[]{\includegraphics[width=0.4\textwidth]{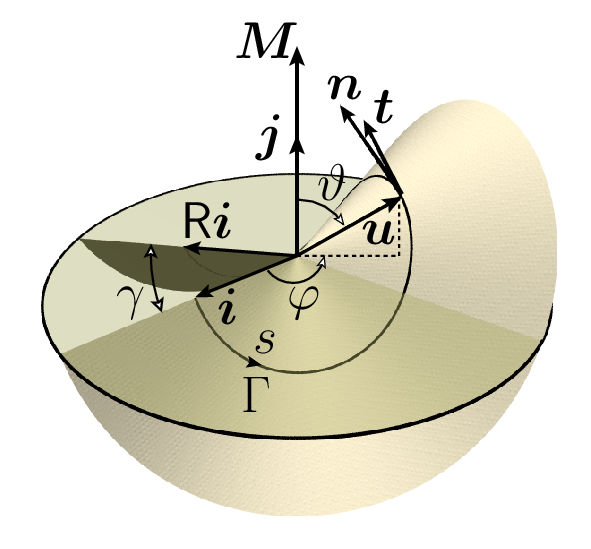}\label{fig:VECM}}
  \hspace*{0.5cm}
  \subfigure[]{\includegraphics[width=0.4\textwidth]{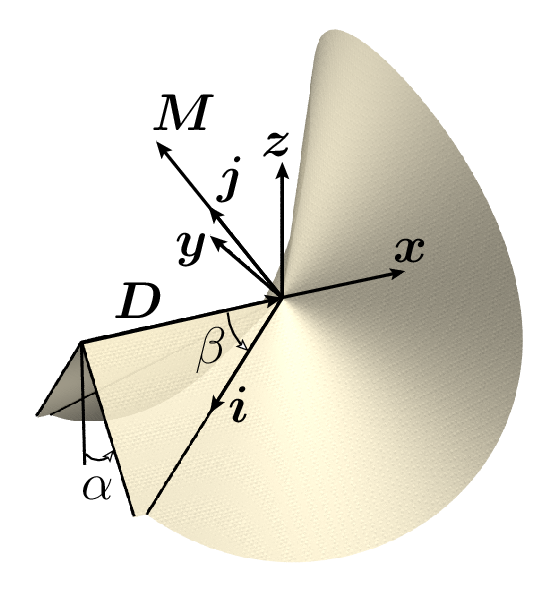}\label{fig:coordsys}}
  \caption{Definition of variables for the surplus cone. The same conventions are used for the deficit defect.}
\label{fig:surpluspart}
\end{figure}
The vector $\VECM_\pm$ is conserved in each of the cones, but will not generally  be aligned with the vertical axis $\VECz$.  Let each $\VECM$ be aligned along a unit vector $\VECj$, with $\VECM=M\VECj$ and $M>0$ (see Fig.~\ref{fig:VECM}). Due to reflection symmetry, $\VECM$  must lie in the mirror plane
$\VECD\VECz$. Let the generators along the boundary of each 
cone point in the directions $\VECi=\VECu|_{s=0}$ and $\sfR\VECi=\VECu|_{s=L_\pm}$, where $\sfR$ represents a reflection in the mirror plane.
For each of the two cones, the vector $\VECi$ will be parameterized by two angles: the angle $\beta\in\{0,\pi\}$ it makes with the dipole vector  
and the angle $\alpha\in\{0,\pi\}$ the $\VECD\VECi$ plane makes with the mirror plane $\VECD\VECz$ (see Fig.~\ref{fig:coordsys}).
When the dipole strength is zero, $\alpha=\pi/2$ and the sheet is flat. 
The angle $\alpha$ must be the same for both cones. Given $\Psi$ and the ratio $D/R$, 
the geometries we consider are completely characterized by $\alpha$, which will be chosen by minimizing the total bending energy of the sheet.

The length $L$ of the curve $\Gamma$ depends on the angles $\beta$ and $\Psi$:
$L_\pm = 2\pi - 2\beta_\pm \pm \se$.
It is evident that $\Psi$ cannot exceed $2\pi$, a bound on the deficit and, thus, the strength of the ``charge-neutral'' dipole. We cannot take the dipole limit $\Psi\to\infty$, $D\to 0$ with $\Psi D = \text{const.}>0$.  There is no exact analogue of an ideal electrostatic point dipole.

\begin{figure}
\centering
  \subfigure[]{\includegraphics[width=0.4\textwidth]{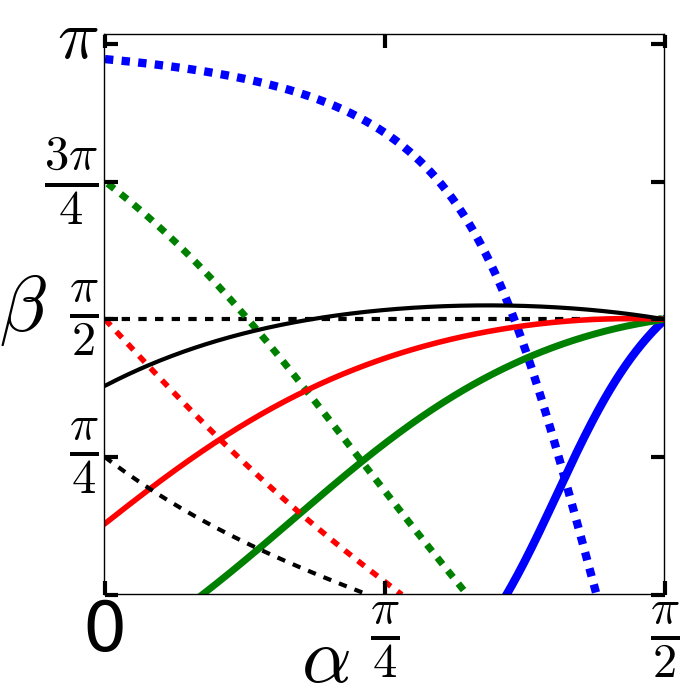}\label{fig:anglebetatorque}}
  \hspace*{0.5cm}
  \subfigure[]{\includegraphics[width=0.4\textwidth]{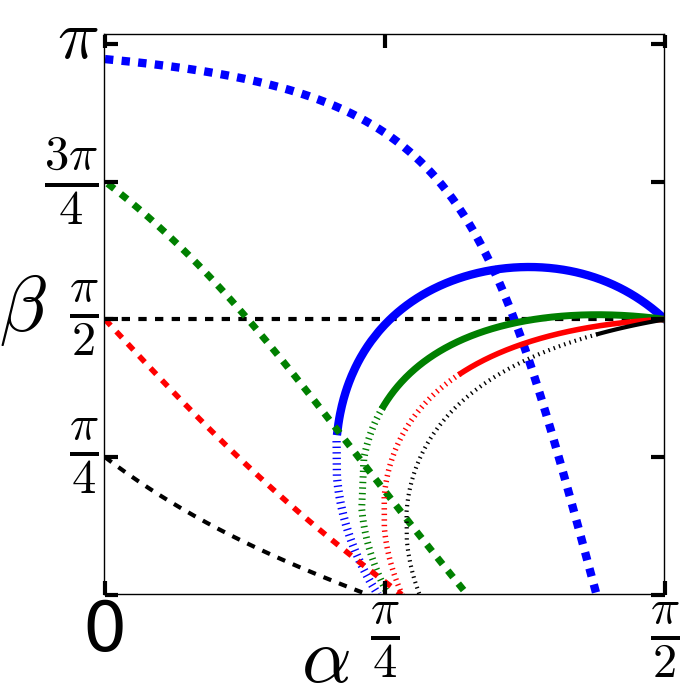}\label{fig:anglebetaforce}}
  \\
  \subfigure[]{\includegraphics[width=0.4\textwidth]{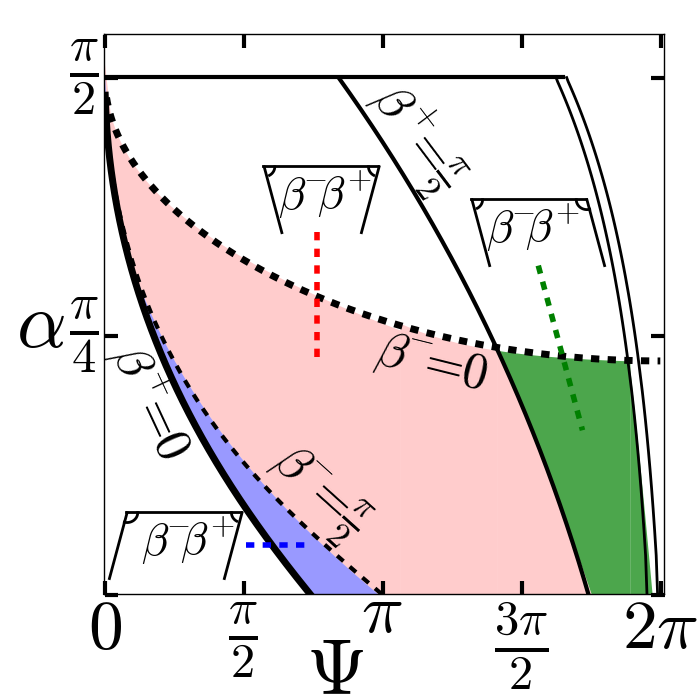}\label{fig:possiblealphastorque}}
  \hspace*{0.5cm}
  \subfigure[]{\includegraphics[width=0.4\textwidth]{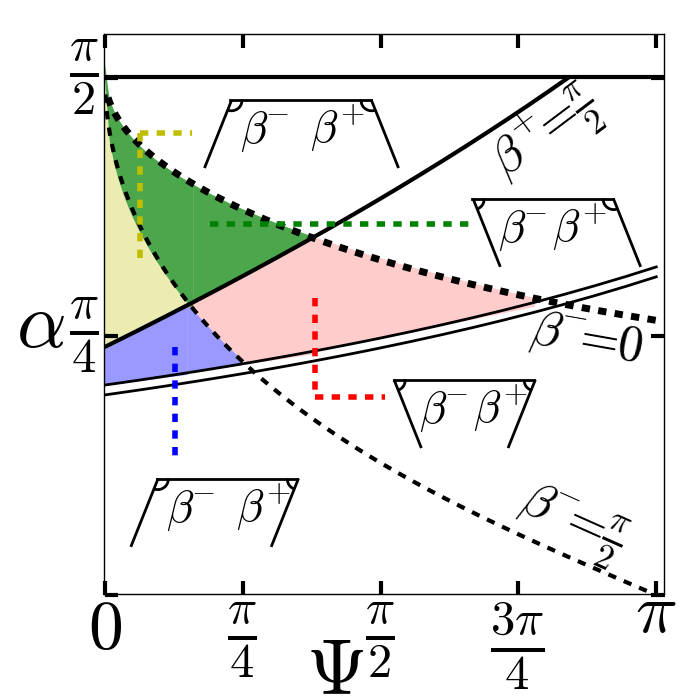}\label{fig:possiblealphasforce}}
  \caption[]{ 
\textit{Upper panel:}
The angle $\beta$ as a function of $\alpha$ for (a) the torque dipole and (b) the force dipole, with strength
$\Psi=\frac{\pi}{18}$ (thick blue curves), $\frac{\pi}{2}$ (green curves), $\pi$ (red curves), and 
$\frac{3\pi}{2}$ (thin black curves). The dashed curves show $\beta_- (\alpha)$, the solid curves $\beta_+ (\alpha)$. Shapes where the surplus part of the force dipole 
intersects with the planar regions correspond to the dotted part of the curves for $\beta_+ (\alpha)$ in (b). For the torque dipole, intersections occur for higher values of $\Psi$ ($\Psi\gtrsim 5.14$).\\
\textit{Lower panel:}
The shaded zone represents those values of $(\Psi,\alpha)$ for which values of $\beta_+$ and $\beta_-$, and thus solutions, may be found for  (c) the torque dipole and (d) the force dipole. 
Different colors represent qualitatively different cases. For example, in the blue region $\beta_- > \frac{\pi}{2}$ and $\beta_+ < \frac{\pi}{2}$. The double curve indicates shapes where the surplus part touches (c) the dipole vector, 
(d) the planar regions of the dipole.
\label{fig:results}
}
\end{figure}
%
As explained in appendix~\ref{app:shapeequation}, we use the projections of $\VECM$ onto the local trihedron $(\VECu,\VECt,\VECn)$ to find the value 
of $\beta_\pm$ for fixed $\Psi$ and $\alpha$: $\VECM\cdot\VECu = M \cos{\vartheta} = \kappa$ provides the polar angle 
$\vartheta$ as a function of the curvature, whereas $\VECM\cdot\VECn = M \,\varphi' \sin^{2}{\vartheta} =  \frac{1}{2}\kappa^{2} - \Cpar$ determines the azimuthal 
angle $\varphi$ in terms of the integrated curvature.\footnote{Instead of examining the problem in its full non-linear glory, one can also confine oneself to 
a harmonic approximation (see appendix~\ref{app:harmonicapproximation}).}  Coupling the surplus and deficit parts in three-dimensional space yields $\beta_\pm$ as a function of $\alpha$.

\section{Results}
The results for the torque dipole are shown in Fig.~\ref{fig:anglebetatorque} 
for several values of $\Psi$. For a fixed $\Psi$ one only finds solutions for $\beta_-$ up to a maximum value of $\alpha$. 
Similarly, there is a minimum value of $\alpha$ 
below which no solution for $\beta_+$ can be found. For $\Psi\to 0$, both of these limits go to $\frac{\pi}{2}$, which corresponds to a planar sheet. 
Solutions for $\alpha>\frac{\pi}{2}$ can be found by inverting the sign of the curvature $\kappa_\pm(s)$. This simply corresponds to a rotation of an equilibrium solution 
around the dipole vector $\VECD$ by an angle of $\pi$. Since such an operation does not change the angle $\beta$, one finds $\beta_\pm(\pi-\alpha)=\beta_\pm(\alpha)$.
This relation is also true for the force dipole. If we again look at $\alpha<\frac{\pi}{2}$, the curves of $\beta_-$ are the same as for the torque dipole.
With the original choice of signs for $\kappa_+(s)$, one only finds solutions for the surplus cone of the force dipole 
when $\alpha>\frac{\pi}{2}$. By rotating about $\VECD$, one can obtain a plot corresponding to Fig.~\ref{fig:anglebetatorque} (see Fig.~\ref{fig:anglebetaforce}):
one again finds a minimum value of $\alpha$  
which is now determined by the requirement that we only want to consider shapes which do not make self-contact. 
In fact, several types of self-contact are possible: $(i)$ the surplus cone can touch the planar regions
of the dipole, $(ii)$ adjacent folds of the surplus cone can touch each other, 
$(iii)$ when $D/R$ is too small, surplus and deficit cones can touch each other in the force dipole configuration. 
Whereas we can exclude case $(iii)$ by assuming that $D/R$ is sufficiently big, we will have to consider $(i)$ and $(ii)$ here. 
Numerically, one finds that the surplus part touches the planar regions before it touches itself (double curve in Figs.~\ref{fig:possiblealphastorque} 
and \ref{fig:possiblealphasforce}).  Results found for the monopole suggest that shapes with self-contact will break symmetry \cite{Stoop2010}.


\begin{figure}
  \subfigure[]{\includegraphics[width=0.47\textwidth]{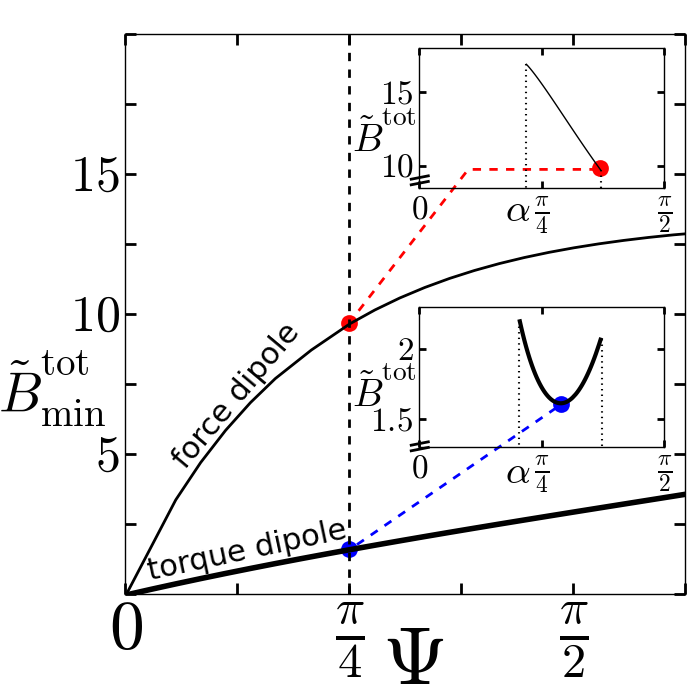}\label{fig:energies}}
  \hspace*{0.5cm}
  \subfigure[]{\includegraphics[width=0.47\textwidth]{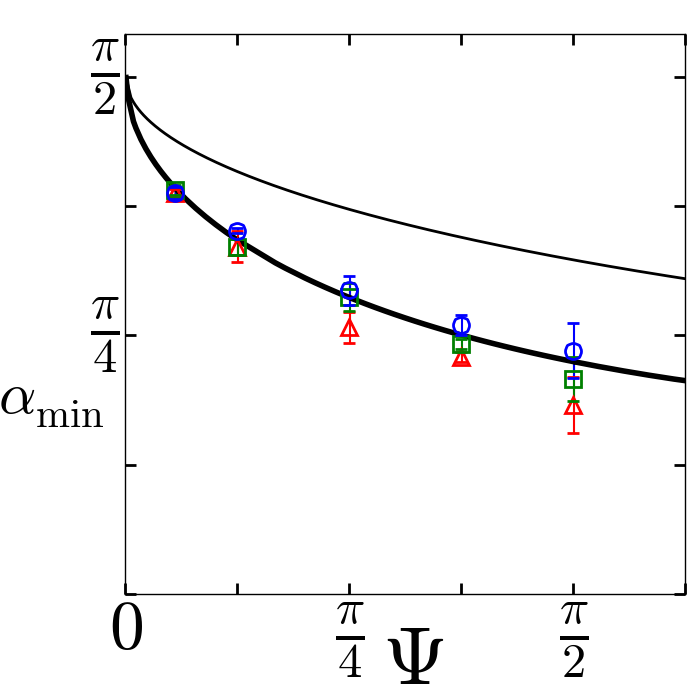}\label{fig:alphamin}}
  \caption{
(a) Minimal bending energy $\tilde{B}^{\text{tot}}_\text{min}$ as a function of $\Psi$ for the torque (thick curve) and the force dipole (thin curve). 
Insets: Total bending energy $\tilde{B}^{\text{tot}}$ of each dipole as a function of $\alpha$ for $\Psi=\frac{\pi}{4}$. 
(b) Dependence of the angle $\alpha_\text{min}$ on $\Psi$ determined semi-analytically (solid curves) for both dipoles, and experimentally for the torque dipole using an A4 sheet of paper with dipole vectors $D=2\,$cm (red triangles), $4\,$cm (green squares), and $6\,$cm (blue circles).  Points and error bars are mean and standard deviation of four measurements.}
  \label{fig:coupledcones}
\end{figure}

We determine the shape of the dipole by combining these isolated solutions.  The angle $\alpha$ has to be the same for both parts. 
In general, the region in the $(\Psi,\alpha)$ space in which a solution can be found is restricted, 
as can also be seen in the lower panel of Fig.~\ref{fig:results}.
To determine which $\alpha$ is predicted by our theoretical model, we minimize the total bending energy of the dipole. We neglect the influence of the width 
of the sheet and the length of the dipole vector $D$, and consider 
$\tilde{B}^\text{tot}=\tilde{B}_{+}+\tilde{B}_{-}$, 
where $\tilde{B}_{\pm} = \int_{0}^{L_\pm} \!\romd s \, \kappa_{\pm}^{2} / 2
= 8 n_\pm^2 \ellipticK{k} (\ellipticE{k} - \ellipticK{k}) / L_\pm$. 
The function $\ellipticE{k}$ is the complete elliptic integral of the second kind \cite{Abramowitz}. The insets of Fig.~\ref{fig:energies} show $\tilde{B}^\text{tot}$ as a 
function of $\alpha$ for $\Psi=\frac{\pi}{4}$. For both dipoles we find one minimum $\tilde{B}^{\text{tot}}_\text{min}$ at an angle $\alpha_\text{min}$ (circular dots in the insets). 
Interestingly, the sum of the conserved torque vectors $\VECM^\text{tot}=\VECM_+ + \VECM_-$ is vertical at this minimum.
The qualitative behaviour of $\tilde{B}^\text{tot}$ does not depend on $\Psi$:
for the torque dipole, the minimum always lies somewhere in the middle of the curve, whereas for the force dipole, it coincides with the upper limit where $\beta_-=0$. 

These results help us to understand the behavior of the paper dipole (see Fig.~\ref{fig:epoles}). We observe that it is simpler to 
convert a force dipole into a torque dipole than \emph{vice versa}. In terms of our theory, this can be explained by the difference in energy of the two states: 
for fixed $\Psi$, the torque dipole always has a lower energy than the force dipole (see Fig.~\ref{fig:energies}). Since both conformations are stable 
with respect to small deformations, there must be an energy barrier between them. 

Moreover, we can now compare the theoretical value of $\alpha_\text{min}$ to experiments. We constructed four paper models of the dipole for each of five strengths $\Psi$ and three dipole lengths $D$. For the force dipole,
our theory predicts that $\alpha_\text{min}$ will be as large as possible.  However, the real physical sheet can also stretch outside of the singularities, and thereby increase the 
dihedral angle even further. Consequently, as mentioned earlier, the real ridge is not entirely straight in the paper model in the force dipole configuration. The angle $\alpha_\text{min}$ seems to be approximately equal to $\frac{\pi}{2}$ close to the ridge, but its value depends on where along the ridge it is measured.  In the torque dipole configuration, $\alpha_\text{min}$ is approximately constant along the ridge. The value of $\alpha_\text{min}$ for each model was obtained in the torque dipole configuration using one of two different methods.  In one, we placed a compass on the ridge and subsequently measured the angle with a protractor.  In the other, we determined $\tan(\alpha_\text{min})$ as given by half of the wingspan of the model divided by the vertical height of the ridge. The results are plotted in Fig.~\ref{fig:alphamin}. One finds that $\alpha_\text{min}$ increases 
with $D$ for fixed $\Psi$, a dependence that our model does not consider. This effect is, however, small compared to the functional dependence of $\alpha_\text{min}$ on the dipole strength $\Psi$.
Our theory explains this functional dependence rather well.


\section{Discussion}
We have found, \emph{via} paper constructions, two stable local minima of bending energy for a dipole in an elastic sheet.  We have assumed that each of these states  can be decomposed into two conical singularities joined by flat parts.  This apparently crude assumption explains the resulting non-trivial shapes surprisingly well. 
The force dipole we have constructed is reminiscent of a well-known fundamental component of crumpling \cite{benpom}, while its torque counterpart resembles a numerically-generated buckled dislocation in a crystalline sheet \cite{SeungNelson88}. 
The rigidity of the torque dipole recommends it as a potentially useful element in lightweight load-bearing structures.   For either construction to be incorporated within a larger system of defects, a closer look at their far-field asymptotic geometry will be necessary.  It is possible that other structures may be induced by the interaction of the constituent monopoles \cite{Witten09}.  The role of this interaction in the mechanical response of the surface is another relatively unexplored topic \cite{Hamm04,Walsh11}, although in this case the distinction between our constructions and the more mobile analogues induced by forcing may be important.


\begin{acknowledgments}
The authors thank Norbert Stoop for helpful comments, and acknowledge the generous hospitality of the Universit\'{e} de Lorraine, Casa F\'{i}sica at UMass Amherst, and UNAM.
JG acknowledges support from DGAPA PAPIIT grant IN114510-3 and CONACyT grant 180901.  
JAH thanks C D Santangelo for support during a postdoctoral appointment through US NSF grants DMR 0846582 and DMR 0820506.
\end{acknowledgments}

\appendix

\section{Determining the curvature\label{app:determinecurvature}}
If we assume $\Cpar<1$ and write $\lambda^{-2}:=1 - \Cpar$, the first integral of the shape equation \eqref{eq:firstintegral} can be integrated:
\begin{equation}
  L =  \int_{\Gamma} \romd s 
    = \int_{\Gamma} \frac{\romd \kappa}{\sqrt{(M^{2} - \Cpar^{2}) - (\frac{1}{4}\kappa^{4} + \lambda^{-2} \kappa^{2})}}
  \, , \quad \text{where } M^{2} - \Cpar^{2}>0 \; .
\end{equation}
Inversion yields
\begin{equation}
  \kappa (s) = \frac{\sqrt{2}\sqrt{-1 + \eta}}{\lambda} \;\Jacobisnadjust{\frac{\sqrt{1+\eta}}{\sqrt{2}\lambda} \, s}{\frac{1-\eta}{1+\eta}}
  \; ,
\end{equation}
where $\eta = \sqrt{1 + (M^{2} - \Cpar^{2}) \lambda^4}\ge 1$. The function $\Jacobisnadjust{s}{k}$ is the sine of the Jacobi
amplitude $\Jacobiamadjust{s}{k}$ with parameter $k$  \cite{Abramowitz}.

The constant $\lambda$ can now be determined by requiring that $\kappa(0)=\kappa(L)=0$.
There is, however, a qualitative difference between the two defects: the curvature of the conical part with the deficit angle 
has one extremum, while the curvature of the part with the surplus angle has three. 
Consequently, one obtains for $n_+=3$ and $n_-=1$:
\begin{equation}
  \lambda_\pm = \frac{\sqrt{1+\eta} \, L_\pm}{2 n_\pm  \sqrt{2} \, \ellipticK{\frac{1-\eta}{1+\eta}}}
  \; .
\end{equation}
The symbol $\ellipticK{k}$ denotes the complete elliptic integral of the first kind \cite{Abramowitz}. 
Without loss of generality we assume that 
the curvature of the deficit part has one maximum and \emph{no} minimum whereas the curvature of the surplus part has 
two minima and one maximum. 
If we define $k:=\frac{1-\eta}{1+\eta}\in [-1,0]$, we obtain:
\begin{equation}
  \kappa_\pm(s) =  \mp \frac{4 n_\pm \sqrt{-k}\, \ellipticK{k}}{L_\pm} \;\Jacobisnadjust{\frac{2 n_\pm \ellipticK{k}}{L_\pm} \, s}{k}
  \; .
  \label{eq:curvature}
\end{equation}
The solutions with the opposite sign, $i.e.$, with the opposite extrema, can be obtained by rotating the resulting shapes 
by an angle of $\pi$ around the dipole vector. 
Since $\lambda^{-2}:=1 - \Cpar$, one has:
\begin{equation}
  \Cpar_{\pm} =   1 - \frac{4 n_\pm^2 (1+k) \, \ellipticK{k}^2}{L_\pm^2}
  \; .
  \label{eq:Cpar}
\end{equation}
A similar calculation can be done for $\Cpar>1$. One finds the same expressions for $\kappa_\pm(s)$ and $\Cpar$, but now $k=\frac{1+\eta}{1-\eta}<-1$. 

Using the quadrature~(\ref{eq:firstintegral}) at $s=0$, $M = \sqrt{\kappa'(0)^2 +\Cpar^2}$, one can determine the absolute value of the torque vector $\VECM$ in terms of 
$k$ and $L_\pm$:
\begin{equation}
  M_\pm =   \sqrt{-\frac{64 n_\pm^4  \, k\, \ellipticK{k}^4}{L_\pm^4} + \left[ 1 - \frac{4 n_\pm^2 (1+k) \, \ellipticK{k}^2}{L_\pm^2} \right]^2}
  \; .
  \label{eq:M}
\end{equation}
The only parameters left undetermined are $k$ and $L$. They are given implicitly by the boundary conditions at $\VECi$ and $\sfR\VECi$ (see next section).


\section{Identifying the two conical shapes \label{app:shapeequation}}
\begin{table}
\centering
\begin{tabular}{|c||c|c|}
  \hline
  $C_+$ & $\beta<\frac{\pi}{2}$ & $\beta>\frac{\pi}{2}$ \\
  \hline\hline
  $\alpha < \frac{\pi}{2}$ & -- & + \\
  \hline
  $\alpha > \frac{\pi}{2}$ & + & -- \\
  \hline
\end{tabular}
  \hspace{1cm}
\begin{tabular}{|c||c|c|c|}
    \hline
  $C_-$ & $\beta<\frac{\pi}{2}$ & $\beta>\frac{\pi}{2}$ \\
  \hline\hline
  $\alpha < \frac{\pi}{2}$ & + & --\\
  \hline
  $\alpha > \frac{\pi}{2}$ & -- & +\\
  \hline
\end{tabular}
   \caption{The sign of $C$  depends on the value of $\alpha$ and $\beta$. \label{tab:C}}
\end{table}
%
\begin{table}
\centering
\begin{tabular}{|c||c|c|c|c|c|}
  \hline
  \raisebox{-0.3cm}{$\phi_+$} &  \multicolumn{2}{c|}{$\beta<\frac{\pi}{2}$} &  \multicolumn{2}{c|}{$\beta>\frac{\pi}{2}$} \\
  & $\phi_+< 0$ & $\phi_+ >0$ & $\phi_+ > 0$ & $\phi_+ <0$ \\
  \hline\hline
  $\alpha < \frac{\pi}{2}$ & n.\ p. & $2\pi - \gamma$ & $2\pi - \gamma$ & self-contact\\
  \hline
  $\alpha > \frac{\pi}{2}$ & self-contact & $\gamma$ & $\gamma$ & n.\ p. \\
  \hline
\end{tabular}
  \hspace{1cm}
\begin{tabular}{|c||c|c|c|c|c|}
    \hline
  \raisebox{-0.3cm}{$\phi_-$} &  \multicolumn{2}{c|}{$\beta<\frac{\pi}{2}$} &  \multicolumn{2}{c|}{$\beta>\frac{\pi}{2}$}  \\
  & $\phi_-< 0$ & $\phi_- >0$ & $\phi_-> 0$ & $\phi_- <0$ \\
  \hline\hline
  $\alpha < \frac{\pi}{2}$ & $\gamma - 2\pi$ & $\gamma$ & $\gamma$ & n.\ p. \\
  \hline
  $\alpha > \frac{\pi}{2}$ & n.\ p. & n.\ f. & n.\ f. & n.\ f.\\ 
  \hline
\end{tabular}
  \caption{The functional dependence of $\phi_\pm = \varphi_\pm(L_\pm)$ on the angle $\gamma$ depends on the sign of $\phi_\pm$ and the 
  value of $\alpha$ and $\beta$ (see text). Shapes with both $C_\pm <0$ and $\phi_\pm< 0$ 
  are not possible (= n.\ p. in the table). For the deficit cone one does not find solutions if $\alpha > \frac{\pi}{2}$  (= n.\ f. in the table).
\label{tab:phi}}
\end{table}
%

To find the value of $\beta$ for fixed angles $\se$ and $\alpha$, we use the projections of $\VECM$ onto the
local trihedron $(\VECu,\VECt,\VECn)$ (see Fig.~3 of the main text).  
The direction of $\VECu$ in Euclidean space is fixed by its azimuthal and polar angles on the unit sphere, 
$\varphi$ and $\vartheta$ respectively. The azimuthal angle is measured with respect to $\VECi$ and the polar 
angle with respect to $\VECj$. Projecting 
\begin{equation}
  \VECM = \left( \frac{\kappa^{2}}{2} - \Cpar \right) \, \VECn
    + \kappa ' \VECt + \kappa \VECu
  \; 
  \label{eq:VECM}
\end{equation}
onto $\VECu$
\begin{equation}
  M (\VECj\cdot\VECu) = M \cos{\vartheta} = \kappa
  \label{eq:projualter}
  \;
\end{equation}
provides the polar angle $\vartheta$ as a function of the curvature  $\kappa$. Eqn.~(\ref{eq:projualter}) places a strong
constraint on the equilibrium shape. In particular it implies that the radial lines where $\kappa$ vanishes lie 
on the plane perpendicular to $\VECM$, \textit{i.e.}, $\VECi\perp\VECj$ and $\sfR\VECi\perp\VECj$. It also implies that 
the curvature $\kappa$ is strictly positive above this plane and strictly negative below it.

Projecting  $\VECM$ onto $\VECt$ reproduces the derivative of
Eqn.~(\ref{eq:projualter}). The remaining projection
\begin{equation}
M (\VECj\cdot\VECn)
    = M \,\varphi' \sin^{2}{\vartheta}
    =  \frac{1}{2}\kappa^{2} - \Cpar 
  \label{eq:projnalter}
  \;
\end{equation}
determines the azimuthal angle $\varphi$ in terms of the integrated curvature
\begin{equation}
  \varphi' =  \frac{1}{M \sin^{2}{\vartheta}} \left( \frac{1}{2}\kappa^{2} - \Cpar \right)
    = \frac{\frac{1}{2}\kappa^{2} - \Cpar}{M (1 - \frac{\kappa^{2}}{M^{2}})}
  = \frac{1}{2} M \left[ \frac{M^{2} - 2\Cpar}{M^{2}}
    \left(\frac{1}{1-M^{-2}\kappa^{2}}\right) - 1\right]
  \; .
  \label{eq:varphidef}
\end{equation}
It is known that \cite{Abramowitz}
\begin{equation}
  \int \frac{\romd s}{1 - b \, \Jacobisnsquare{c \,s}{k}}
  = \frac{1}{c} \ellipticpiadjust{b}{\Jacobiamadjust{c \, s}{k}}{k}
  \; ,
  \label{eq:ellipticintegralthirdkind}
\end{equation}
where $\Pi$ is the elliptic integral of the third kind. Thus, with Eqs.~(\ref{eq:varphidef}) and (\ref{eq:ellipticintegralthirdkind}),
\begin{eqnarray}
  \varphi_\pm (s)  & = &  \int_{0}^{s} \romd t \, \varphi'(t)
  \\
   & = & \frac{M_\pm^{2} - 2 \Cpar_\pm}{2 M_\pm}
    \frac{L_\pm}{2 n_\pm \ellipticK{k}}
    \ellipticpiadjust{\frac{16 n_\pm^2 (-k)\ellipticK{k}^{2}}{M_\pm^{2} L_\pm^{2}}}
    {\Jacobiamadjust{\frac{2 n_\pm \ellipticK{k} \, s}{L_\pm}}{k}}{k}
    - \frac{M_\pm}{2} s
  \; . \;\;\;\;\;\;\;\;\;
  \label{eq:varphiofs}
\end{eqnarray}

Introduce a Cartesian chart on the surplus cone, adapted to the geometry of the dipole. The unit vector $\VECx$ is parallel to 
the dipole vector $\VECD$ and points away from the ridge (see Fig.~3 of the main text). 
This, together with a vector $\VECy$ and the vertical vector $\VECz$, defines a right-handed basis.
Likewise, one can define a different Cartesian chart on the deficit cone with a unit vector $\VECx$ anti-parallel to $\VECD$. 

At $s=0$ the trihedron $(\VECu,\VECt,\VECn)$ is then given by:
\begin{equation}
  \VECi =  - \cos\beta\, {\VECx} + \sin \beta\, \VECi_\perp \, , 
  \quad   
  \VECt_{s=0} = \sin\beta\, {\VECx} + \cos \beta\, \VECi_\perp \, , 
  \quad 
  \VECn_{s=0} = -\cos{\alpha} \, \VECy + \sin{\alpha} \, \VECz
  \; ,
\end{equation}
where ${\VECi}_\perp = -\sin\alpha \,\VECy -\cos \alpha \,\VECz$. 
The unit vector $\VECj$ in the direction of $\VECM$ can also be expressed in terms
of the Cartesian basis $(\VECx,\VECy,\VECz)$:
\begin{equation}
 \VECj \sim \frac{\VECi\times \sfR\VECi}{|\VECi\times \sfR\VECi|} = \frac{\cos\alpha \sin\beta \, \VECx -
  \cos\beta \, \VECz} {\sqrt{ 1- \sin^2 \alpha \sin^2 \beta}} 
  \; .
\end{equation}

The projection of $\VECM=M\VECj$ onto $\VECt$ has to be equal to $\kappa'_\pm$ (see Eqn.~(\ref{eq:VECM})). With the definition of 
signs in Eq.~(\ref{eq:curvature}), one obtains the sign of $\VECj$ as follows:
\begin{equation}
\label{eq:Jdef}
 \VECj_\pm^{\alpha<\frac{\pi}{2}}= \mp \frac{\VECi\times \sfR\VECi}{|\VECi\times \sfR\VECi|} \,, \quad \text{and} \quad 
   \VECj_\pm^{\alpha>\frac{\pi}{2}} = -  \VECj_\pm^{\alpha<\frac{\pi}{2}}
\end{equation}
since we have defined $M>0$.
The projection at $s=0$ thus gives us for all $\alpha\in\{0,\pi\}$:
\begin{equation}
  \frac{M_\pm \left|\cos{\alpha}\right| }{\sqrt{1-\sin^2{\alpha} \sin^2{\beta_\pm} }} = \frac{8 n_\pm^2 \sqrt{-k}\, \ellipticK{k}^2}{L_\pm^2}
  \; .
  \label{eq:projt}
\end{equation}
Projecting $\VECM$ onto $\VECn$ at $s=0$ yields:
\begin{equation}
  \Cpar_\pm^{\alpha<\frac{\pi}{2}} = \mp\frac{M_\pm \, \sin{\alpha} \cos{\beta_\pm} }{\sqrt{1-\sin^2{\alpha} \sin^2{\beta_\pm} }}
  \,, \quad \text{and} \quad 
  \Cpar_\pm^{\alpha>\frac{\pi}{2}} = -  \Cpar_\pm^{\alpha<\frac{\pi}{2}}
  \; .
  \label{eq:projn}
\end{equation}
The sign of $\Cpar$ depends on the value of $\alpha$ and $\beta$ (see Tab.~\ref{tab:C}): 
$\operatorname{sign}{C_\pm}=\mp \operatorname{sign}{\left[\left( \frac{\pi}{2} - \alpha \right) \left( \frac{\pi}{2} - \beta \right) \right]}$. 

Finally, we have to connect $\varphi(s)$ to the extrinsic geometry. The range of the azimuthal angle $\phi=\varphi (L)$ on 
the $\VECi\sfR\VECi$ plane depends on the angle $\gamma$ between $\VECi$ and $\sfR\VECi$ in a nontrivial way (see Tab.~\ref{tab:phi}). 
$\gamma$ is given by
\begin{equation}
  \cos{\gamma} = \VECi \cdot \sfR \VECi = \cos^2{\beta} + \cos{(2\alpha)}\sin^2{\beta} 
  \; , \quad\text{where } \gamma\in\{0,\pi\}
  \; ,
\end{equation}
whereas one finds the following expression for $\phi$ with the help of Eqn.~(\ref{eq:varphiofs}):
\begin{equation}
    \phi_\pm=\varphi_\pm (L_\pm)  = L_\pm \left( \frac{M_\pm^{2} - 2 \Cpar_\pm}{2 \ellipticK{k} M_\pm}
    \ellipticpicompleteadjust{\frac{16 n_\pm^2 (-k)\ellipticK{k}^{2}}{M_\pm^{2} L_\pm^{2}}}{k}
    - \frac{M_\pm}{2} \right)
  \; ,
  \label{eq:phi}
\end{equation}
where we have used that $\Jacobiamadjust{2 n_\pm \ellipticK{k}}{k}=n_\pm \pi$ for $k<1$.
The value of $\phi$ can in principle be positive or negative. For $\phi >0$, 
one either has $\phi = \gamma$ if $\VECj$ and $\VECi\times \sfR\VECi $ are parallel, or $\phi = 2\pi - \gamma$ 
if the two vectors are anti-parallel (see Eqn.~(\ref{eq:Jdef})). When $\phi <0$, one finds $\phi = \gamma - 2\pi$ if $\VECj$ and $\VECi\times \sfR\VECi $ are parallel, and  $\phi = -\gamma$ if they are anti-parallel. 
Using Eqn.~(\ref{eq:projt}) and the condition on $\phi$ (see Eq.~(\ref{eq:phi}) and Tab.~\ref{tab:phi}) one can now calculate $\beta_\pm$ as a function of $\alpha$. 

It is instructive to discuss the sign of $\phi$ in connection with the sign of $\Cpar$ (see Tab.~\ref{tab:C}). 
If $\Cpar$ is negative, Eqn.~(\ref{eq:varphidef}) 
directly tells us that $\varphi'(s) >0$ for all $s\in\{0,L_\pm\}$ and consequently $\phi>0$ in this case. This implies that both $\Cpar<0$ and $\phi<0$ 
are not possible (= n.\ p. in Tab.~\ref{tab:phi}). For $\Cpar$ positive, the situation is more complicated.  $\varphi'$ at $s=0$ is now always 
smaller than zero, but changes sign along $\Gamma$ if $\Cpar$ is not too big. To determine $\Cpar^{\text{cr}}$, where $\phi$ changes sign we can, for instance, 
consider the projections of $\VECM$ at $s=L/2$ where the curvature $\kappa$ is maximal: $\phi_\pm$ changes sign when $\vartheta(L/2)=0$, $i.e.$, 
when $M^{\text{cr}}\stackrel{(\ref{eq:projualter})}{=}\kappa(L/2)$ and 
$2\Cpar^{\text{cr}}\stackrel{(\ref{eq:projnalter})}{=}\kappa^2(L/2)=(M^{\text{cr}})^2$.


\section{Harmonic approximation \label{app:harmonicapproximation}}

Let us suppose that $\Psi /2\pi \ll 1$, and expand the curvature $\kappa(s)$, as well as
the constants $\Cpar$ and $M$, in powers of $\epsilon =\sqrt{-k}$:
\begin{equation}
  \kappa = \kappa_1 + \kappa_3  + \cdots\,;\quad M= M_0+ M_2 + \cdots\,;\quad 
    C = C_0 + C_2 + \cdots
  \; .
\end{equation}
For $\sqrt{-k}\to 0$, one has $\ellipticK{0}=\frac{\pi}{2}$ and $\Jacobisnadjust{m \, s}{0} = \sin{(m\, s)}$. 
The equilibrium states are described by small oscillations about an equator, where $\kappa=0$.  In the harmonic
approximation of the quadrature~(\ref{eq:firstintegral}) about $\kappa=0$,
\begin{equation}
\label{eq:linCEELsphere}
  M_0^2 - \Cpar_0^2 + 2 M_0 M_2 - 2 \Cpar_0 \Cpar_2 = (\kappa_1')^2 + (1 - \Cpar_0) \kappa_1^2
  \; .
\end{equation}
At lowest order, the arc length coincides with the azimuthal angle,
$\varphi$. The curvature along a loop, vanishing at
$\varphi=0$ and at $L_0$, is given by
\begin{equation}
\label{eq:linear}
  | \kappa_1 | \approx A_1\sin \left(\frac{\pi n \varphi}{L_0}\right)
  \; ,
\end{equation}
where the amplitude $A_1$ is a constant. From Eq.~(\ref{eq:curvature}) we infer:  
$A_1 = \frac{2\pi n}{L^0} \sqrt{-k}$.

The quadrature implies the two zeroth order identities 
\begin{equation}
\label{eq:sig0}
\Cpar_0 = 1 - \left(\frac{n \pi}{L_0}\right)^2
\, , \quad
M_0^2 = \Cpar_0^2
\; ,
\end{equation}
where $M_0$ is positive, as one can show by expanding Eq.~(\ref{eq:M}). 
The sign of $\Cpar_0$ depends on whether $n \pi$ is bigger or smaller 
than $L_0$ (see below). 

Eq.~(\ref{eq:linCEELsphere}) also implies
\begin{equation}
2 (M_0 M_2 - \Cpar_0 \Cpar_2) = A_1^2 (1 - \Cpar_0)
\end{equation}
and thus
\begin{equation}
\label{eq:kappa1sM} 
  A_1^2  =   \frac{- 2 \Cpar_0}{1 - \Cpar_0} [ - M_2 \operatorname{sign}{(\Cpar_0)} + \Cpar_2 ]
  \; .
\end{equation}
This is a constraint on a combination of the second order corrections $M_2$ and $\Cpar_2$. As a 
check of consistency, one can expand the nonlinear expressions for $\Cpar$ and $M$, Eqs.~(\ref{eq:Cpar}) and (\ref{eq:M}), in $\sqrt{-k}$ and 
show that they indeed obey Eqs.~(\ref{eq:sig0}) and (\ref{eq:kappa1sM}).

The amplitude $A_1$ depends on the ratio of arc length to polar angle over the trajectory
on the unit sphere.  One has the exact formula
\begin{equation}
L_\pm = 2\pi - 2\beta \pm \Psi  = \int_0^{|\phi|}  d\varphi \,
\left[\left(\frac{d \vartheta} {d \varphi}\right) ^2 + \sin^
2\vartheta \right]^{1/2}
\; .
\end{equation} 
where $|\phi|=2\pi - \gamma$ if $\beta<\frac{\pi}{2}$ or $|\phi|=\gamma$  if $\beta>\frac{\pi}{2}$.  
Using Eq.~(\ref{eq:projualter}) and the harmonic approximation for $\kappa$ given by
Eq.~(\ref{eq:linear}), one obtains
\begin{eqnarray}
\sin^2{\vartheta}  & \approx & 1 - \frac{A_1^2}{M_0^2} \sin^2{\left(\sqrt{1-\Cpar_0} \,\varphi\right)} 
\,, \quad \mbox{and} \quad 
\\
\left( \frac{d \vartheta} {d \varphi} \right)^2 & \approx & \frac{1}{M_0^2} \, \left(\frac{d \kappa_1 } {d
\varphi} \right)^2 = \frac{(1-\Cpar_0) \, A_1^2}{M_0^2} \, \cos^2{\left(\sqrt{1-\Cpar_0}\,\varphi\right)}
\; .
\end{eqnarray}
Thus, in the quadratic approximation,
\begin{equation}
\label{eq:A1L} 
2\pi - 2\beta \pm \Psi \approx |\phi| \left(1 -  \frac{A_1^2}{4 C_0}\right) \,.
\end{equation}
Note that $|\phi|=L_0 + \mathcal{O}(\sqrt{-k})$. 
Together Eqs. (\ref{eq:kappa1sM}) and (\ref{eq:A1L}) imply the identity
\begin{equation}
\label{mCpar2} 
-M_2 \operatorname{sign}{\Cpar_0} + \Cpar_2 = 2 (1 - \Cpar_0) (\Gamma-1)\,,
\end{equation}
where $\Gamma= (2\pi - 2\beta \pm \Psi )/|\phi|$.

This is all we need to satisfy the boundary conditions at lowest non-trivial 
order \cite{JemalPablo2012}.
The remaining boundary condition~(\ref{eq:projn}) only involves $\Cpar$ and $M$ in
the combination $-M_2 \operatorname{sign}{(\Cpar_0)} + \Cpar_2 $ at second order, or
\begin{equation}
\frac{\Cpar}{M} \approx \frac{\Cpar_0}{M_0} + \frac{M_0 C_2 - C_0 M_2}{M_0^2} 
= \operatorname{sign}{(\Cpar_0)} \left[ 1 + \frac{\Cpar_2 - M_2 \operatorname{sign}{(\Cpar_0)} }{\Cpar_0} \right]
\; .
\end{equation}
Using Eq.~(\ref{mCpar2}) this reads
\begin{equation}
  \frac{\Cpar}{M} \approx \operatorname{sign}{(\Cpar_0)} \left[ 1 + \frac{2 (1-\Cpar_0)}{\Cpar_0} (\Gamma-1)\right]
  \; .
\end{equation}

We now approximate the right-hand side. To this end, we need an expression for the angle $\gamma$. 
When $\Psi$ is small,  $\alpha \approx \pi/2$, and it is possible to approximate
\begin{equation}
\cos \gamma \approx \cos{(2 \beta)} + 2 \left(\frac{\pi}{2}-\alpha\right)^2 \sin^2{\beta}
  \; ,
\end{equation}
or, equivalently,
\begin{eqnarray}
\label{eq:gammalinbetasm90}
\gamma & = &  2 \beta - \Delta\gamma \, , \qquad\; \quad \text{if } \beta < \frac{\pi}{2} \qquad \text{and}
\\
\label{eq:gammalinbetagr90}
\gamma & = &  2 \pi - 2 \beta + \Delta\gamma \, , \quad \text{if } \beta > \frac{\pi}{2} \; .
\end{eqnarray}
where we have defined $\Delta\gamma= \left(\frac{\pi}{2}-\alpha\right)^2 \tan\beta$.

For both $\beta < \frac{\pi}{2}$ and $\beta > \frac{\pi}{2}$, we thus have $|\phi|=2\pi-2 \beta + \Delta\gamma$ and, 
\begin{equation}
  \Gamma -1 \approx \frac{1}{2\pi - 2\beta} \left[ \pm \Psi - \Delta\gamma \right]
  \; .
\end{equation}
Furthermore, using Eq.~(\ref{eq:sig0}) we obtain
\begin{equation}
-\frac{\Cpar_0}{1-\Cpar_0} = 1- \left(\frac{|\phi|}{n\pi}\right)^2 \approx 1 -\left(\frac{2\pi -2\beta
}{n\pi}\right)^2 - 2 \Delta \gamma \left(\frac{2\pi
-2\beta}{n^2\pi^2}\right) 
  \; ;
\end{equation}
the correction in $\gamma$ is irrelevant at this order. One thus has
\begin{equation}
\frac{2 (1-\Cpar_0)}{\Cpar_0} (\Gamma-1) \approx
 \frac{2}{2\pi - 2\beta}
\Bigg[ 1  - \left(\frac{2\pi -2\beta
}{n\pi}\right)^2 \Bigg]^{-1}
\left( \mp \Psi  + \Delta\gamma \right)
  \; .
\end{equation}

When $\alpha\approx \frac{\pi}{2}$, the trigonometric function on the right-hand side
of Eq.~(\ref{eq:projn}) reduces to
\begin{equation}
  \frac{ \sin\alpha \cos\beta}{\sqrt{ 1- \sin^2 \alpha \sin^2 \beta}} 
  \approx \operatorname{sign}{\left(\frac{\pi}{2}-\beta\right)} 
    \left[ 1 - \frac{1}{2} \left(\frac{\pi}{2}-\alpha\right)^2 \sec^2{\beta} \right]
  \; .
\end{equation}

To formulate the boundary condition (\ref{eq:projn}) in the correct manner, we have to take a closer look and 
determine the signs of the right-hand and the left-hand side of the equation for the different cases.

We know that $L_0$ has to be smaller than $2\pi$. Since $n_+=3$ for the surplus part, we directly obtain from Eq.~(\ref{eq:sig0}) 
that $\Cpar_{+,0}\le-\frac{5}{4}$ is always negative and $M_0=-\Cpar_{+,0}$. The (nonlinear) $\Cpar_+$ is thus also negative 
for small $\Psi$; one only finds two types of solution (see Tab.~\ref{tab:C}): ($\alpha<\frac{\pi}{2}$, $\beta_+<\frac{\pi}{2}$, $L_0=2\pi-\gamma$) and 
($\alpha>\frac{\pi}{2}$, $\beta_+>\frac{\pi}{2}$, $L_0=\gamma$).

Consequently, the boundary condition (\ref{eq:projn}) reads  
\begin{equation}
  \frac{2}{2\pi - 2\beta} \Bigg[ 1 - \left(\frac{2\pi -2\beta}{3 \pi}\right)^2 \Bigg]^{-1}
  \left( - \Psi  +  \left(\frac{\pi}{2}-\alpha\right)^2 \tan\beta\right) =
  -\frac{1}{2}\left(\frac{\pi}{2}-\alpha\right)^2 \sec^2\beta
\label{eq:BCsurplus}
\; .
\end{equation}
for all $\beta_+\in\{0,\pi\}$.

For the deficit part $n_-=1$. Since for $\beta_- < \frac{\pi}{2}$ we always have $L_0 = 2\pi - \gamma > \pi$, we find that $\Cpar_{-,0}$ is positive.
For $\beta_- > \frac{\pi}{2}$ we always have $L_0 = \gamma < \pi$ and consequently $\Cpar_{-,0} < 0$. 
Both types of solution with $\alpha < \frac{\pi}{2}$ are thus allowed (see Tab.~\ref{tab:C} for the sign of $\Cpar_-$).
However, $\alpha > \frac{\pi}{2}$ is not allowed. 

The boundary condition (\ref{eq:projn}) for $\beta_-\in\{0,\pi\}$ becomes: 
\begin{equation}
 \frac{2}{2\pi - 2\beta} \Bigg[ 1 - \left(\frac{2\pi -2\beta}{\pi}\right)^2 \Bigg]^{-1}
  \left( \Psi  + \left(\frac{\pi}{2}-\alpha\right)^2 \tan\beta\right) =
  -\frac{1}{2}\left(\frac{\pi}{2}-\alpha\right)^2 \sec^2\beta
\label{eq:BCdeficit}
\; .
\end{equation}

Eqs.~(\ref{eq:BCsurplus}) and (\ref{eq:BCdeficit}) can be reorganized in the following form:
\begin{equation}
\label{eq:phidelalpha2}
\Psi= \frac{1}{2} \, f_\pm(\beta) \, \left(\frac{\pi}{2}-\alpha\right)^2
  \; ,
\end{equation}
where
\begin{eqnarray}
f_\pm(\beta) &=& \pm\frac{1}{2}\, (2\pi - 2\beta)
\sec^2\beta  \left[1 -\left(\frac{2\pi -2\beta}{n_\pm\pi}\right)^2\right]  \pm 2\tan\beta \nonumber\\
&=& \pm \frac{1}{2}\, \sec^2 \beta \left[
 (2\pi - 2\beta)
\left[1 -\left(\frac{2\pi -2\beta}{n_\pm \pi}\right)^2\right]  + 2\sin{2 \beta} \right]
  \; .
\end{eqnarray}
For a nonvanishing $\Psi$, Eq.~(\ref{eq:phidelalpha2}) implies that $\alpha=\frac{\pi}{2}$ is only possible  
for values of $\beta_\pm$ at which $f_\pm$ diverges. The function $f_+$ of the surplus part diverges at $\beta_+=\frac{\pi}{2}$ 
whereas the one for the deficit part is a monotonically decreasing function of $\beta_-$ with no divergences. Thus, $\left.\beta_+\right|_{\alpha = \frac{\pi}{2}}=\frac{\pi}{2}$ for any
$\Psi$, whereas $\left.\beta_-\right|_{\alpha = \frac{\pi}{2}}=0$ only if $\Psi\to 0$. The same behavior is found in the nonlinear regime.
Moreover, the harmonic approximation and the numerical results of the full nonlinear equations coincide in the limit of vanishing $\Psi$.

The angle $\alpha$ has to be the same for the two parts. For fixed $\Psi / 2\pi \ll 1$, this requirement, together with Eq.~(\ref{eq:phidelalpha2}), implies that $f_+(\beta_+) = f_-(\beta_-)$ in the harmonic approximation.


\end{document}